\documentclass[aps, twocolumn,amsmath, ammsymb]{revtex4-1}
\usepackage{epsfig}

\usepackage{bbm}
\usepackage{amsmath}
\usepackage{amssymb}
\usepackage{txfonts}
\usepackage{graphicx}
\usepackage{dcolumn}
\usepackage[usenames]{color}
\begin{document}

\preprint{}

\title{Edge states and phase diagram for graphene under polarized light}
\author{Yi-Xiang Wang}
\affiliation{School of Science, Jiangnan University, Wuxi 214122, China.}
\author{Fuxiang Li}
\affiliation{Center for Nonlinear Studies and Theoretical Division, Los Alamos National Laboratory, Los Alamos, NM 87545 USA.}
\date{\today}

\begin{abstract}
In this work, we investigate the topological phase transitions in graphene under the modulation of circularly polarized light, by analyzing the changes of edge states and its topological structures.  A full phase diagram, with up to ten different topological phases, is presented in the parameter space spanned by the driving frequency and light strength.  We find that the high-Chern number behavior is very common in the driven system.  While the one-photon resonance can create the chiral edge states in the $\pi-$gap, the two-photon resonance will induce the counter-propagating edge modes in the zero-energy gap. When the driving light strength is strong, the number and even the chirality of the edge states may change in the $\pi-$gap.  The robustness of the edge states to disorder potential are also examined.  We close by discussing the feasibility of experimental proposals.
\end{abstract}


\maketitle

\section{\label{sec:level1} Introduction}

One of the central themes in recent condensed matter physics is searching for new topological states, e.g. the topological insulators (TIs) \cite{M.Z.Hasan,X.L.Qi}, which are insulating in the bulk but support topologically protected metallic states at the boundary.  This area of studies has becoming more interesting thanks to the introduction of periodical light to the existing materials.  Specifically, by applying periodic, short laser pulse to topological, or even trivial, materials, the driven systems are expected to undergo topological phase transitions, leading to the so-called Floquet TIs \cite{N.H.Lindner,B.Dora}.  Floquet TIs are intrinsically different from the traditional TIs since they are in non-equilibrium rather than in equilibrium.  Experimentally, the Floquet states have been reported to be observed on the surface of Bi$_2$Se$_3$ \cite{Y.H.Wang} as well as in a hexagonal lattice made in a photonic crystal \cite{M.C.Rechtsman}.

Since the application of periodic light to the electron system breaks the time-reversal symmetry (TRS), the resulted Floquet TI belongs to the quantum anomalous Hall (QAH) class, according to the ten-fold classification of topological matter by dimensionality and certain discrete symmetries \cite{A.P.Schnyder}.  However, in contrast to the static TI characterized by Chern number, which, according to the bulk-boundary correspondence, is directly related to the total number of edge modes in the gap \cite{Y.Hatsugai}, for the Floquet TI, the topological states are intrinsically dynamical and the correspondence between the bulk and edge states is more subtle.  For example, for a two-band two dimensional system under driving light, in addition to the energy gap located near zero energy, another inequivalent gap would appear at the boundary of first Brillouin zone (BZ) in frequency domain, and in some cases, it would support chiral edge modes despite the fact that the Chern numbers associated with both bands are zero.  Therefore, the Chern number characterization in static system  is insufficient to describe its topological structure. One approach to this problem is proposed by Rudner \emph{et. al} \cite{M.S.Rudner}, in which a new topological invariant is constructed that yields the correct edge state structure in the driven case.

For graphene in particular, it has been proposed that the application of polarized light can turn the trivial equilibrium bands into the nontrivial Floquet TI and different dynamical generalizations of static topological phases can be achieved \cite{T.Kitagawa,J.I.Inoue,A.Kundu,Z.H.Gu,A.G.Leon,G.Usaj,P.Titum,L.E.F.Torres,A.Quelle,M.A.Sentef, Perez2014, Perez2015}. However, a global phase diagram that can provide us with the information about the specific phase for certain driving protocol has not yet been reported. In Ref. \cite{A.G.Leon}, a phase diagram in the parameter space spanned  by  the field amplitude and phase polarization has been reported for the modulated honeycomb lattices. Since the driving frequency plays a more important role in determining the topological property of the system, here we try to derive the phase diagram containing the driving frequency by using the Floquet theorem.  The phase diagram will be calculated by analyzing the edge states and the topological structures.  We will mainly focus on the drivings of intermediate frequency that is comparable to the bandwidth of the system.

As the band overlap may occur in the whole momentum region, compared with many previous studies which focus on the low-energy approximation, here the exact diagonalization calculation is more reliable.  Our work shows that phase diagram of the driven graphene ribbon contains up to ten different topological phases, in the parameter space spanned by the driving frequency and light strength.  Among them, exist the phases with high-Chern number (Chern number larger than 1)  as well as the phases with  counter-propagating edge modes in the gap.  In each phase, the topological structure and its physical origin are analyzed.  We also examine the robustness of the edge states to the inevitable disorder potential.  Finally, the possible experimental proposals of the driven graphene results and the detection method are discussed.

\section{\label{sec:level1} Model and Method}

We start from the noninteracting electron system defined on a honeycomb lattice and coupled to a circularly polarized light.  For the polarized light, only the effect of its electric field part is considered, while that of the magnetic field part is neglected.  The electric field is assumed to be in-plane and spatially homogeneous.  We can describe the polarized light by the vector potential ${\bf A}(\tau)=A_0(\text{cos}\omega\tau,\text{sin}\omega\tau)$, where $A_0$ is the amplitude and $\omega=\frac{2\pi}{T}$ is the frequency.  In general, the interaction between the light and electrons can be included in the hopping integrals as $t_{ij}(\tau)=te^{iA_{ij}(\tau)}$ where $t$ is the nearest-neighboring hopping strength and $A_{ij}(\tau)=\frac{e}{\hbar}{\bf A}(\tau)\cdot({\bf r}_i-{\bf r}_j)$ is the phase factor which strongly depends on the hopping direction.

As the whole system is time-periodic, the powerful Floquet theorem \cite{H.Sambe,J.H.Shirley} can be applied.  The two-component wavefunction $\psi_{\bf k}(\tau)$ encoding the time-dependent amplitudes on each sublattice can be written as $\psi_{\bf k}(\tau)=e^{-i\epsilon_{\bf k}\tau/\hbar}u_{\bf k}(\tau)$, where $\epsilon_{\bf k}$ is the quasienergy and $u_{\bf k}(\tau)=u_{\bf k}(\tau+T)$ is the periodic function.  Substituting $\psi_{\bf k}(\tau)$ into the time-dependent Schr\"{o}dinger equation $i\hbar\partial_\tau\psi_{\bf k}(\tau)=H(\tau)\psi_{\bf k}(\tau)$, the latter will be reduced to the following eigenvalue problem:
\begin{eqnarray}
H_{F}({\bf k},\tau)u_{\bf k}(\tau)=\epsilon_{\bf k}u_{\bf k}(\tau),
\end{eqnarray}
where $H_F({\bf k},\tau)=H({\bf k})-i\hbar\frac{\partial}{\partial\tau}$ is the Floquet Hamiltonian.  We can expand $u_{\bf k}(\tau)$ into the Fourier series as $u_{\bf k}(\tau)=\sum_{n=-\infty}^{\infty}u_{n{\bf k}}e^{in\omega \tau}$, in which $u_{n{\bf k}}=(u_{n{\bf k}A},u_{n{\bf k}B})^T$ denotes the $n$th Floquet mode corresponding to the quasienergy $\epsilon_{\bf k}$.  Using the formula of $e^{iz\text{cos}\omega\tau}=\sum_{n=-\infty}^{\infty}J_n(z)e^{in(\omega \tau+\frac{\pi}{2})}$, with $J_n(x)$ the n-th order Bessel function, Eq. (1) can be written as the time-independent form:
\begin{eqnarray}
&&n\omega u_{n{\bf k}A}-t\sum_m J_m(\alpha)f_{m{\bf k}} u_{n-m{\bf k}B}=\epsilon_{\bf k}
u_{n{\bf k}A},  \\
&&-t\sum_m J_m(\alpha)g_{m{\bf k}}  u_{n+m{\bf k}A}+n\omega u_{n{\bf k}B}=\epsilon_{\bf k} u_{n{\bf k}B},
\end{eqnarray}
where $t$ is the nearest-neighbor (NN) hopping integral and $\alpha=\frac{eA_0a_0}{c}$ is the dimensionless strength parameter for light.  The functions $f_{m{\bf k}}=e^{i{\bf k}\cdot{\bf a}_1}+e^{i({\bf k}\cdot{\bf a}_2+\frac{4m\pi}{3})}+e^{i({\bf k}\cdot{\bf a}_3+\frac{2m\pi}{3})}$ and $g_{m{\bf k}}=e^{-i({\bf k}\cdot{\bf a}_1+m\pi)}
+e^{-i({\bf k}\cdot{\bf a}_2-\frac{m\pi}{3})}+e^{-i({\bf k}\cdot{\bf a}_3+\frac{m\pi}{3})}$, where ${\bf a}_l=a_0(\text{cos}\beta_l,\text{sin}\beta_l)$ are the NN vectors for a honeycomb lattice and $\beta_l=\frac{(4l-1)\pi}{6}$.  Note that $(f_{m{\bf k}})^*\neq g_{m{\bf k}}$.  Therefore, one has the infinite dimensional eigenvalue problem in the direct product Floquet space of ${\mathcal R}\bigotimes{\mathcal T}$ \cite{H.Sambe}, with $\mathcal R$ being the usual Hilbert space and ${\mathcal T}$ the space of periodic functions spanned by the function of $\text{exp}(in\omega t)$.  To solve it effectively, we follow the standard procedure to truncate the Floquet Hamiltonian $H_F$ into a finite dimensional matrix with $-M\leq n\leq M$ \cite{M.S.Rudner}.  This is justified because the $n-$th Floquet component $u_{n{\bf k}}$ decays rapidly with $|n|$ beyond a finite range in the frequency space. In actual calculations, when choosing $M=5$ the accurate results can be obtained already.

In a periodically driven system, the temporal periodicity gives rise to the periodicity in frequency space,  that is, the quasienergy $\epsilon_{\bf k}$ is equivalent to $\epsilon_{\bf k}+p\omega$ for any integer $p$. Therefore, we often set the quasienergy to live within the quasienergy Brillouin zone (QBZ) as $\epsilon_{\bf k}\in[-\frac{\omega}{2},\frac{\omega}{2}]$. In the system, we should distinguish two kinds of gaps \cite{A.G.Leon}: the zero-energy gap $\Delta_0$ between the conduction and valence bands within $n=0$ Floquet bands as well as the $\pi-$gap $\Delta_\pi$ separating the conduction and valence bands of the neighboring Floquet bands. Physically, the zero-energy gap is caused by the first-order perturbation contribution of virtual processes involving the photon absorption and emission in which the two processes do not commute and thus impart an effective mass to the low-energy Floquet states, while the $\pi-$gap is produced by the avoided crossings between the $n=0$ and $n=\pm1$ Floquet bands \cite{T.Oka,L.E.F.Torres}.

As there are two inequivalent gaps which will not close simultaneously, there may be topologically distinguished states but with the same Chern number.  When the edges are present in the system, we consider the upper edge of the system and denote the number of chiral edge modes propagating in the $\hat x(-\hat x)$ direction by $n_u^+(n_u^-)$, where the index $u=0,\pi$ characterizes the zero-energy gap and the $\pi-$gap, respectively. Rudner \textit{et al.} showed the net chiralities of the edge modes as $C_0=n_0^+-n_0^-$ and $C_{\pi}=n_{\pi}^+-n_{\pi}^-$.  Further the Chern number of the conduction band in the Floquet system can be obtained from the net chiralities as $C=C_0-C_\pi$ \cite{M.S.Rudner}.

In the numerical calculations, the number of $n_u^+$ and $n_u^-$ can be obtained exactly by counting the edge states from the ribbon dispersion according to the bulk-edge correspondence \cite{Wang2, M.Ezawa}.  In the quasienergy spectrum,  besides the bulk bands, there exist  edge states, which always appear in pairs.  Each pair of chiral edge states contributes one unit to $n_u^{+(-)}$ in the corresponding gap.  More precisely, to evaluate $n_u^{+(-)}$, we need to take into account their locations (upper or bottom edges), which can be determined by analyzing their wave functions, as well as their directions (right or left) of propagation, which can be obtained from the sign of the group velocity $\frac{d\epsilon}{dk_x}$.

\section{\label{sec:level1} Main Results}

\textit{Phase diagram. --} The graphene system under the modulation of light can exhibit diverse topological edge states, whose behavior strongly depends on the driving frequency $\omega$ and the light strength parameter $\alpha$.  In the high-frequency regime $\omega\gg W$ where $W=6t$ is the bandwidth, the Floquet Hamiltonian in Eqs. (2) and (3) is approximately block diagonal.  The neighboring Floquet bands are decoupled and well separated by a shift of $\omega$.  When the frequency decreases and becomes comparable to the bandwidth $\omega\sim W$, a hierarchy of band-crossings will appear,  giving rise to energy gaps that may hold edge states.   As a result, the number of edge states will change and the topological phase transition happens.  Physically, the band-crossings can be understood as one-photon or multi-photon resonance. When the frequency further decreases and is much smaller than the bandwidth $\omega\ll W$, such band-crossings become serious. The gap between the neighboring Floquet bands, as well as the edge states within it, thus become obscure \cite{A.G.Leon}.  Because of this reason, we mainly focus on the case of intermediate-frequency $\omega\sim W$, even though it is expected that graphene under low-frequency driving would host new states that are absent in the high-frequency region.  As for the parameter $\alpha$, it renormalizes the hopping integrals, and thus leads to the change of the bandwidth $W$ and correspondingly to the change of the topological structure of the system.

\begin{figure}
\includegraphics[width=9.2cm]{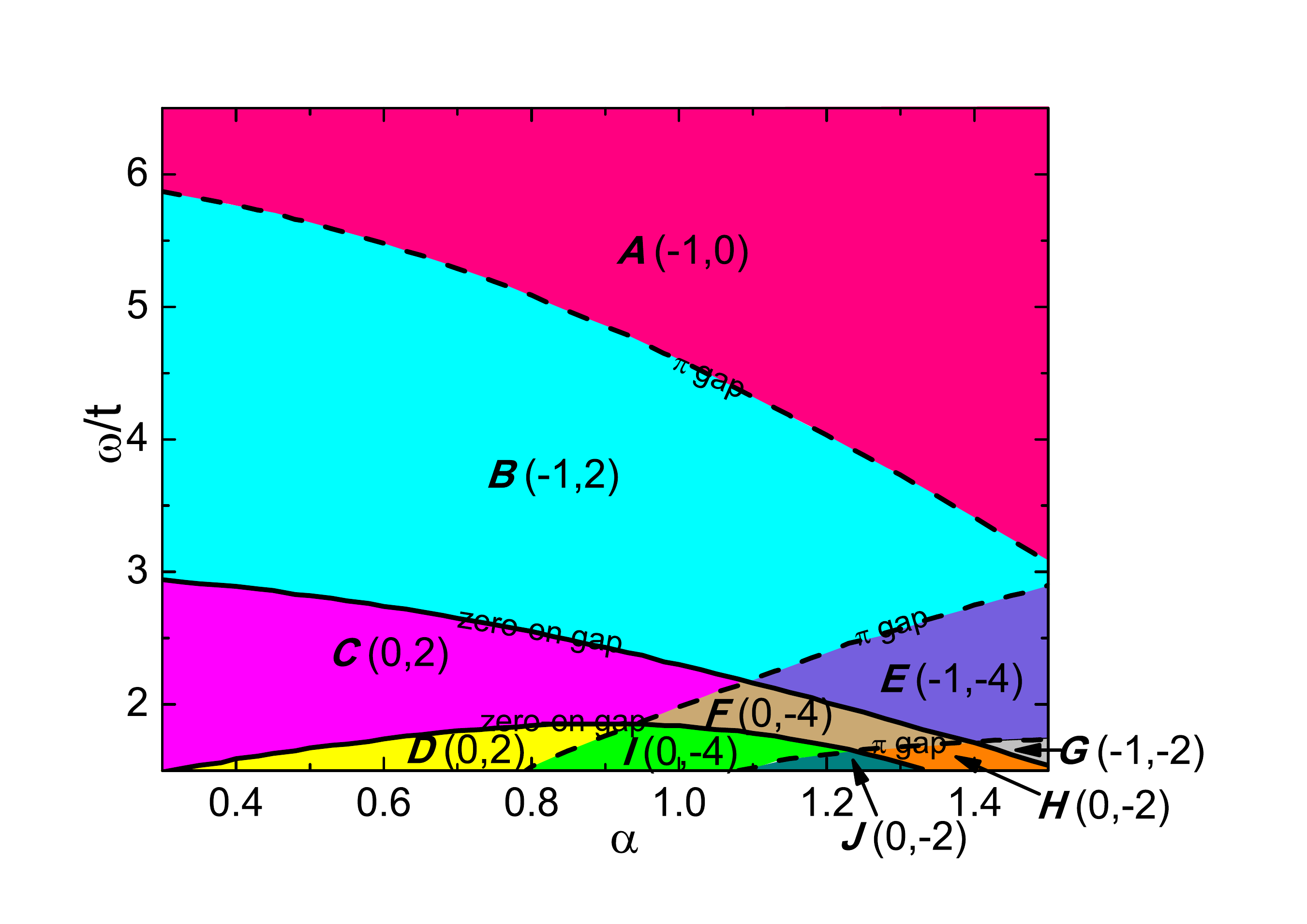}
\caption{(Color online) Phase diagram of periodically driven graphene in the parameter space spanned by $\alpha$ and $\omega$.  Each phase ($A$ to $J$) are characterized by $(C_0,C_\pi)$.  We have labeled the phase boundary as solid lines when zero-energy gap is closed and as dotted lines when $\pi-$gap is closed.}
\end{figure}

\begin{table}
\begin{tabular}{c|c|c|c|c|c|c|c|c|c|c}\hline
& \emph{A}& \emph{B}& \emph{C}& \emph{D}& \emph{E}& \emph{F}& \emph{G}& \emph{H}& \emph{I}& \emph{J} \\ \hline
$(n_0^+,n_0^-)$& (0,1)& (0,1)& (1,1)& (2,2)& (0,1)& (1,1)& (0,1)& (1,1) &(2,2) &(2,2)  \\ \hline
$(n_\pi^+,n_\pi^-)$& (0,0)& (2,0)& (2,0)& (2,0)& (0,4)& (0,4)& (0,2)& (0,2)& (0,4)& (0,2)  \\ \hline \hline
$C_0$, $C_\pi$& -1,0& -1,2& 0,2& 0,2& -1,-4& 0,-4& -1,-2& 0,-2& 0,-4& 0,-2  \\ \hline \hline
$C$& -1& -3& -2& -2& 3& 4& 1& 2& 4& 2  \\ \hline
\end{tabular}
\caption{The topological characteristics of each phase from $A$ to $J$, where ${(n_u^+,n_u^-)}$ give the number of chiral modes within different gaps, $C_u=n_u^+-n_u^-$ denotes the net chirality and $C=C_0-C_\pi$ is the Chern number of the Floquet system.}
\end{table}

The distinct topological phases hosted by periodically driven graphene can be classified according to the net chiralities of the bulk bands $(C_0,C_\pi)$ and the characteristics of the edge states in the quasienergy spectrum.  The resulting phase diagram in terms of the driving parameters $\alpha$ and $\omega$ is shown in Fig. 1.   Similar diagrams are presented  in Ref.~\cite{Perez2015}, where the phase transitions in $0$ gap and $\pi$ gap are separately discussed. We can see that the system undergoes versatile phase transitions as a function of $\alpha$ and $\omega$.  For the range of parameters shown, there are five possible phase transition boundaries which  divide the parameter space into ten phases labeled as $A$ to $J$.  All phases have well-defined quasienergy bands: both the zero-energy gap and $\pi-$gap are well formed.  Here the neighboring phases are always separated by a transition boundary, where the band gap closes at certain value of $k_x$ and the band inversion occurs.  At the phase boundaries, either the zero-energy gap or $\pi-$gap is closed, which are denoted, respectively, by solid lines or dashed lines.

To better illustrate these phases, in Table I, the topological characteristics of each phase are summarized, including the number of chiral modes ${(n_u^+,n_u^-)}$, the net chirality $C_u$ within different gaps and the Chern number $C$ of the Floquet system.  It shows that in these phases except phase $A$, there are more than one pair of edge states in the zero-energy gap or $\pi-$gap, indicating the high-Chern number behavior is very common in the Floquet system.  For the chiralities of these edge states, we find that the edge states in the $\pi-$gap always have the same chiralities while in the zero-energy gap, the chiralities may be different (see, phase $C$, $F$ and $H$).  Across the phase transition boundaries along which the zero-energy gap or $\pi$-gap close, the Chern number $|C|$ is changed, respectively, by 1 and 2.

It should be noted that for phases $C$ and $D$, although they own the same net chirality and the same Chern number value, they have different topological structures as the number of chiral modes in the zero-energy gap are different, which can be seen in Fig. 3(c) and (d) below.  The arguments are the same for phases $F$ and $I$ as well as $H$ and $J$.  This suggests the Chern number alone is not enough to describe the number of edge states.  We need to take into account of the symmetry and the topological structure of the Floquet Hamiltonian to understand the topological properties of the driven system.

\begin{figure*}
\centering
\includegraphics[width=18cm]{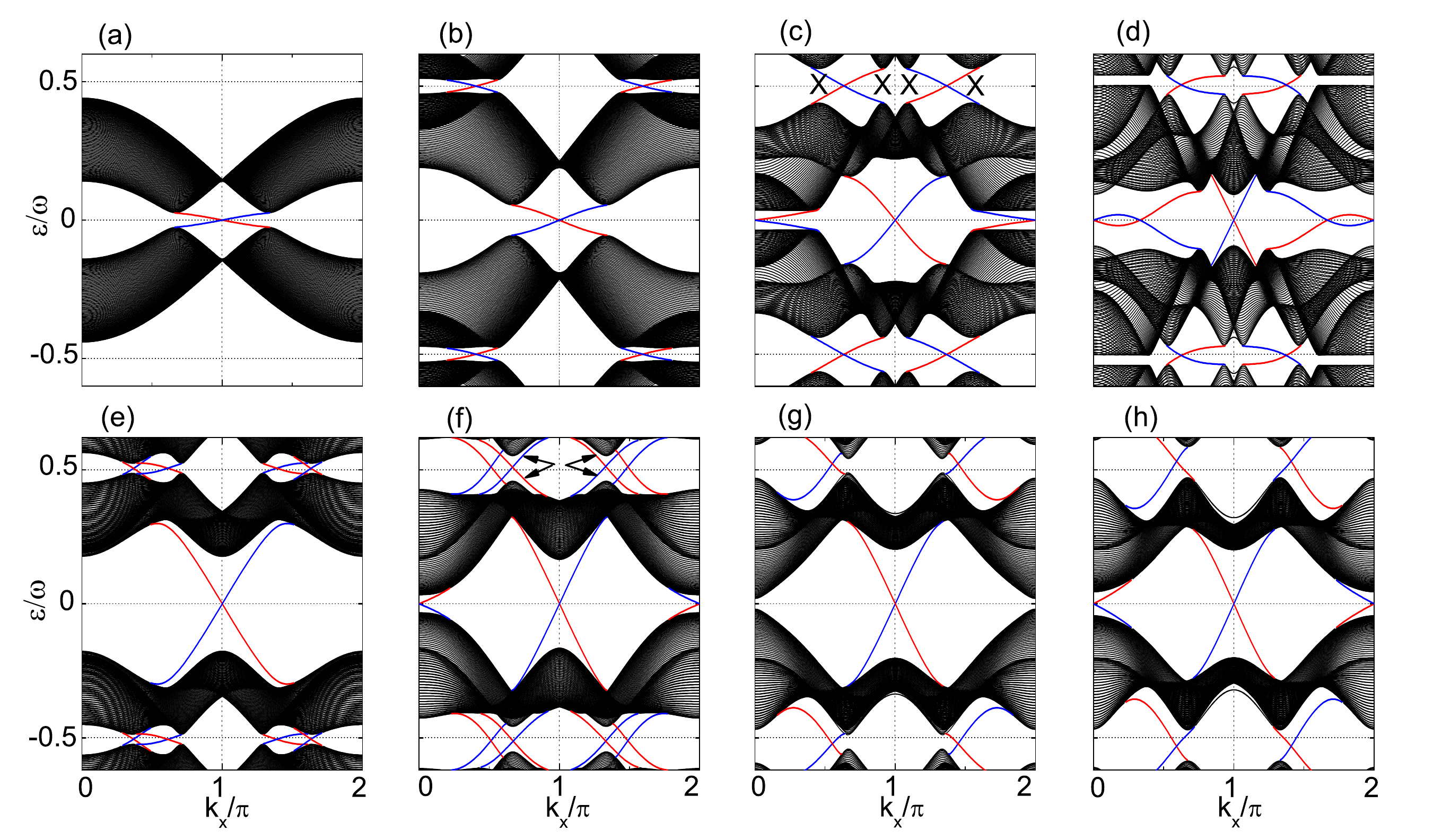}
\caption{(Color online) Quasienergy spectra of a finite ribbon (in the $y$ direction) of the periodically driven graphene system.  The eight panels correspond to phase $A$-$H$, as shown in Fig. 1, respectively.  From (a) to (h), the parameters of the driving $(\alpha,\omega)$ are taken as $(0.7,6)$, $(0.7,4)$, $(0.7,2.25)$, $(0.7,1.52)$, $(1.3,2.3)$, $(1.3,1.8)$, $(1.45,1.55)$, $(1.45,1.65)$.  Edge states localized on the upper and bottom edge are shown in red and blue, respectively.  }
\end{figure*}

\textit{Ribbon dispersion.--} Next in Fig. 2, the representative spectra of phases $A$ to $H$ are shown for a ribbon structure with 200 lattice in the $y$ direction and infinitely long in the $x$ direction.  Due to the periodicity of the Floquet system, we only plot the quasienergy in the range of $(-0.65\omega,0.65\omega)$.  It shows clearly that in all these phases, there exist a pair of chiral edge states crossing at the Dirac point $k_x=\pi$ in the zero-energy gap.  Such edge states are the same for all different values of photon frequencies and light strength. They come from the linearity of the graphene bands around the Dirac point and have no relevance from the photon resonances.

For the ribbon structure, the quasi-energy spectra evidently exhibit several nice symmetries of the Floquet Hamiltonian.  First, the Floquet system has the symmetry of spatial inversion and is invariant under the operation of $k_x\rightarrow-k_x$ and $y\rightarrow L_y-y$.  Thus an edge state solution $\epsilon(k_x)$ implies the existence of another edge state solution $\epsilon(-k_x)$ (or $\epsilon(2\pi-k_x)$) with the same quasienergy but localized at the opposite edge.  Therefore, the edge states at $k_x\neq0,\pi$ always come in pairs.  Second, the Floquet system exhibits the particle-hole symmetry (PHS): if $\epsilon(k_x)$ is a quasienergy eigenvalue, another solution of $-\epsilon(k_x)$ must exist correspondingly. In addition, the PHS combined with the periodicity of the Floquet Hamiltonian lead to the symmetry around $\epsilon=\frac{\omega}{2}$.  In the following, we will discuss these phases one by one.

(1) In phase $A$, i.e. in the high-frequency regime, the electron cannot directly absorb photons due to the off-resonant light.  In $\pi-$gap, as there exist no edge states, it is topologically trivial.  The effective Hamiltonian can take the same form as the static Hamiltonian, only with the hopping integrals being replaced by the effective ones $t_{ij}\rightarrow tJ_0(\alpha)$.  Here the renormalized hopping integrals to the lowest-order do not depend on the hopping directions.  Therefore the driven system can be well compared with the static one except for the renormalized hopping integrals. Early studies about the Floquet TI were mainly focused on this phase \cite{T.Kitagawa,J.I.Inoue,Y.X.Wang}.

(2) In phase $B$, there are two pairs of edge states in the $\pi-$gap.  We can see the conduction band of $n=0$ Floquet band overlap with the valence band of $n=1$ Floquet replica, leading to the band inversion around the $\pi-$gap.  At $\epsilon=\frac{\omega}{2}$, the energy difference between the valence and conduction bands is precisely $\omega$ \cite{A.Quelleb}, which allows for the one-photon resonance.  Note the chiralities of edge states here are different from those in the zero-energy gap.  The transport properties about the dc conductance and quantum Hall response have been investigated in this phase \cite{L.E.F.Torres}, where the transport signatures were suggested to be dominated by the time-averaged density of states.

(3) In phase $C$, inside the zero-energy gap, it shows that besides the edge states at $k_x=\pi$, there appear additional ones at $k_x=0$.  More importantly, the chiralities of these edge states are opposite with those at $k_x=\pi$, i.e. they propagate along the opposite directions at the same edge. Therefore, the states at $k_x=0$ are called the counter-propagating edge modes.  The corresponding topological numbers give $n_0^+=n_0^-=1$.  This is in sharp contrast with the static TIs or the QH effect where the edge states in the same gap must own the same chirality.  Compared with the previous studies, where the counter-propagating edge modes are predicted to appear in the $\pi-$gap of the periodically driven QH system (either the periodically kicked \cite{M.Lababidi} or harmonically driven \cite{Z.Y.Zhou} Hofstadter model ), here our study shows that such modes can also appear in the zero-energy gap of the periodically driven graphene.

\begin{figure*}
\centering
\includegraphics[width=18cm]{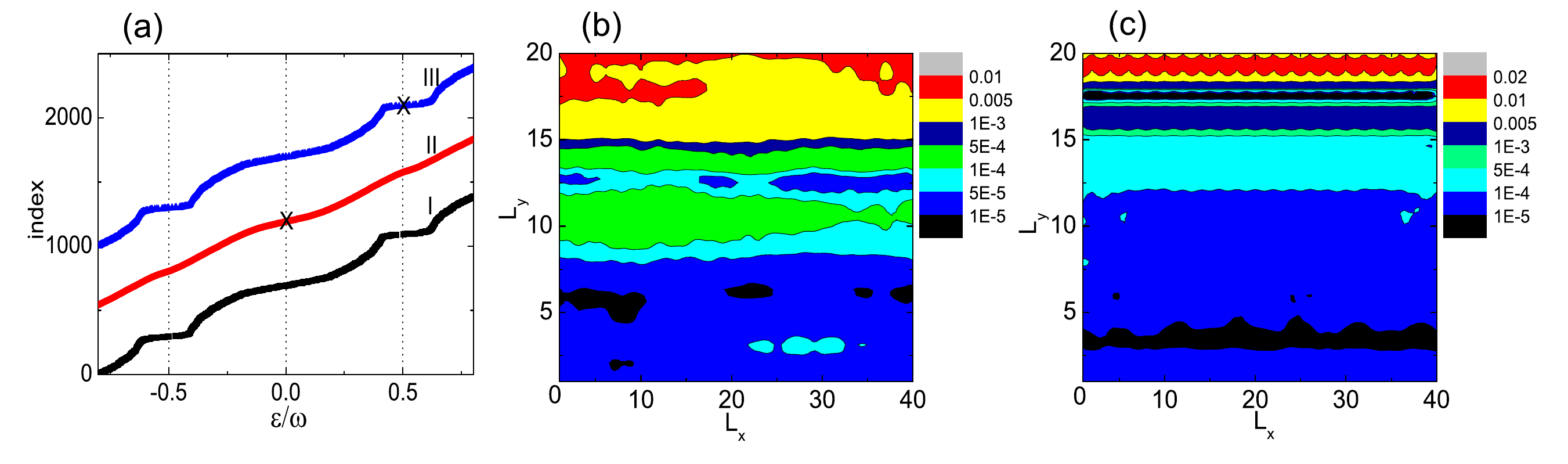}
\caption{(Color online) (a) The quasienergy eigenvalues of a periodically driven graphene on a finite $40\times20$ lattice for $\alpha=1.3$ and $\omega=1.8$, corresponding to phase $F$ in Fig. 2.  The horizontal axis is the index of the eigenvalues. The curve I is the spectrum without perturbation, while II and III are for case (1) and (2), respectively. They have been shifted vertically from each other for clarity.  (b),(c) Contour plot of the wavefunction in the Floquet system when disorder exist.  The parameters used for (b) and (c) correspond to the crosses in line II and III in (a), respectively. }
\end{figure*}

Evidently in Fig. 2(c), to avoid the band-crossings between the valence band of $n=1$ Floquet band and the conduction band of $n=-1$ Floquet band, the band inversion happens around $\epsilon=0$ which creates the counter-propagating edge modes.  At $\epsilon=0$, the energy difference between the conduction band and the valence band is $2\omega$ \cite{A.Quelleb}, so the edge states can be regarded as being induced by the two-photon resonances.  In the framework of the low-energy effective theory, Quelle \textit{et al.} recongnized the significance of the band inversions and referred the edge states appearing at $k_x=0$ as a realization of the Bernevig-Hughes-Zhang model of the HgTe/CdTe quantum well \cite{A.Quelle}.

(4) Phase $D$ can be obtained from phase C by decreasing the frequency.  Intrinsically it has same topological structure as phase C. However, within the zero-energy gap,   the original linear structure of the edge states in phase C evolves into a helical structure, that is, the modes around $k_x=0$ cross each other for three times.  This means that the propagation direction of the original edge states at a given edge changes. Correspondingly the topological numbers give $n_0^+=2$, $n_0^-=2$.  However, the net chiralities remain the same as in phase $C$.  The helical structure in the low-frequency region can be attributed to the nonlinearity of the bands near $k_x=\pi$.

(5) Phase $F$ can be treated as one obtained from phase by increasing the driving magnitude. The difference with phase C is the appearance of four, instead of two, pairs of edge states within the $\pi-$gap. Moreover, these four pair of edge states have the same chirality, but opposite to those in phase $C$. The topological number $n_\pi^+$ vanishes while $n_\pi^-=4$.  This topological phase transition is accompanied by the change of Chern number of conduction band from $-2$ to $4$.

(6) Phase $H$ is obtained from phase $F$ by further increase of the driving magnitude. Within the $\pi-$gap, exist two pairs of edge states with the same chirality as in phase $F$. Correspondingly, the Chern number of the conduction band changes from $4$ to $2$.
From the phase diagram in Fig. 1, it shows only when the light strength $\alpha$ is large that the topological structures that are similar to phase $F$ and $H$ may happen in the $\pi-$gap.  This is because the structures of the Floquet bands are changed (not only becoming flat) due to the strong light field.  As a result, the Chern number will change in each Floquet band and the edge states with different number and chirality will appear.

Here we have studied different topological phases of the periodically driven graphene and found several new phases, from $D$ through $H$, which exhibit novel topological structure that have not been discussed before.  Our analysis suggest that the one-photon resonances can create the edge states in the $\pi-$ gap, while the two-photon resonances will induce the counter-propagating edge states in the zero-energy gap.

\textit{Robustness of the edge states.--} As the driven topological phases belong to the QAH class \cite{A.P.Schnyder}, the edge states should not be affected by the disorder potential inevitably presenting in real materials.  Here we try to numerically check the robustness of the edge states by solving the quasienergy of periodically driven graphene.  The size of the system is set as $L_x\times L_y$.  We choose the periodic boundary condition in the $x-$direction and open boundary condition in the $y-$direction.  Two kinds of static perturbations are considered: (1) random onsite potential $U_i\in[-\frac{t}{2},\frac{t}{2}]$; (2) exceedingly strong onsite potential $U\sim10^3 t$ on the edge (at position $(20,1)$).  In Ref.~\cite{Perez2014}, the robustness was also tested against vacancies included randomly with a very low density.

In Fig. 3(a), we plot the quasienergy spectra of the driven system with the above perturbations included.  For comparison, the quasienergy spectra without disorder is also plotted as curve I.  For curve II, it shows due to the scattering of the impurity potentials, more bulk states are scattered into the zero-energy gap and $\pi-$gap, so the curve becomes smooth.  While for curve III, the strong onsite potential shows minor influence on the low-energy spectra.   The results demonstrate the robustness of the edge states, similar to the static ones. Further in Fig. 3(b) and (c), we plot the contour of the wavefunction $\psi(x,y,t=0)$ corresponding to quasienergy $\epsilon=0$ and $\epsilon=\frac{\omega}{2}$.  We note the main contribution to the wavefunction comes from the $n=0$ Floquet component, indicating the correctness of the cutoff of the photon number in the calculation.  For the given quasienergies, the wavefunctions are both localized on the upper edge.  Compared with the unperturbed case, the presence of disorder distorts the wavefunction but does not break the property of edge states.

\section{\label{sec:level1} Discussions and Summaries}

We discuss the feasibility of our results in experiment. The periodically driven graphene system exhibiting the topological nontrivial phases have been proven to be realized in different physical setups.  In real graphene material the relevant hopping parameter is $t=2.7$eV and the NN bond length $a_0=1.4{\AA}$, which requires rather strong driven light field with the frequency in the THz region and the amplitude $A_0\sim 10^{-2}$V$\cdot$s/m.  There are also proposals with artificial honeycomb structure where the bands can also be described by the tight-binding model, such as the photonic crystals \cite{M.C.Rechtsman} with extended helical wave guides in the third spatial dimension as well as the shaken optical lattices \cite{G.Jotzu,P.Hauke}.  In addition, the microwave honeycomb crystals \cite{M.Bellec} have also been put forward, where the hopping parameter is of the order of a few MHz \cite{M.Bellec} while the driving frequency can be varied up to GHz \cite{S.Gehler}, which makes the experiment more easily to operate.

The Floquet spectrum and the gaps due to the photon resonance in the driven system can be detected by using the ARPES technique, as recently demonstrated on the surface of 3D TI \cite{Y.H.Wang}.  The edge modes in the nontrivial gap can be detected by measuring  the spectral function $\rho(k_x,\omega)$ \cite{F.Li, F.Li2}, which is well captured by momentum-resolved radio-frequency spectroscopy \cite{J.T.Stewart}.  On the other hand, the transport properties of electrons \cite{L.E.F.Torres} may provide the direct signal of the nontrivial phase in the driven system. However, there are less studies in this topic \cite{L.E.F.Torres} and further investigations are needed in the future.

In summary, we have studied the driven graphene under the circularly polarized light and  obtained a global phase diagram as a function of the frequency and strength of the light.  We analyze the topological structures of different phases in detail, especially the the high-Chern number behavior and the counter-propagating edge modes.  We suggest such counter-propagating modes may exist only in the driven system.  In addition, we demonstrate the robustness of the edge modes.  Our work may deepen the understanding of the driven non-equilibrium system and help us to search the new topological states of matter.

\section{\label{sec:level1} Acknowledgements}

This work is supported by Natural Science Foundation of Jiangsu Province, China under grant No. BK20140129.


\begin{references}

\bibitem{M.Z.Hasan} M. Z. Hasan, C.L. Kane, Rev. Mod. Phys. {\bf82}, 3045 (2010).

\bibitem{X.L.Qi} X.L. Qi, S.C. Zhang, Rev. Mod. Phys. {\bf83}, 1057 (2011).

\bibitem{N.H.Lindner} N. H. Lindner, G. Refael, and V. Galitski, Nat. Phys. {\bf7}, 490 (2011).

\bibitem{B.Dora} B. Dora, J. Cayssol, F. Simon, and R. Moessner, Phys. Rev. Lett. {\bf108}, 056602 (2012).

\bibitem{Y.H.Wang} Y. H. Wang, H. Steinberg, P. Jarillo-Herrero, and N. Gedik, Science {\bf342}, 453 (2013).

\bibitem{M.C.Rechtsman} M. C. Rechtsman, J. M. Zeuner, Y. Plotnik, Y. Lumer, D. Podolsky, F. Dreisow, S. Nolte, M. Segev, and A. Szameit, Nature (London), {\bf496}, 196 (2013).

\bibitem{A.P.Schnyder} A. P. Schnyder, S. Ryu, A. Furusaki, and A. W. W. Ludwig, Phys. Rev. B {\bf78}, 195125 (2008).

\bibitem{Y.Hatsugai} Y. Hatsugai, Phys. Rev. Lett. {\bf71}, 3697 (1993).

\bibitem{M.S.Rudner} M. S. Rudner, N. H. Lindner, E. Berg, and M. Levin, Phys. Rev. X {\bf3}, 031005 (2013).

\bibitem{T.Kitagawa} T. Kitagawa, T. Oka, A. Brataas, L. Fu, and E. Demler, Phys. Rev. B {\bf84}, 235108 (2011).

\bibitem{Z.H.Gu} Z. H. Gu, H. A. Fertig, D. P. Arovas, and A. Auerbach, Phys. Rev. Lett. {\bf107}, 216601 (2011).

\bibitem{J.I.Inoue} J. I. Inoue and A. Tanaka, Phys. Rev. Lett. {\bf105}, 017401 (2010).

\bibitem{G.Usaj} G. Usaj, P. M. P. Piskunow, L. E. F. torres, and C. A. Balseiro, Phys. Rev. B {\bf90}, 115423 (2014).

\bibitem{P.Titum} P. Titum, N. H. Lindner, M. C. Rechtsman, and G. Refael, Phys. Rev. Lett. {\bf114}, 056801 (2015).

\bibitem{A.Kundu} A. Kundu, H. A. Fertig, and B. Seradjeh, Phys. Rev. Lett. {\bf113}, 236803 (2014).

\bibitem{A.G.Leon} A. G. Leon, P. Delpalce, and G. Platero, Phys. Rev. B {\bf89}, 205408 (2014).

\bibitem{L.E.F.Torres} L. E. Foa Torres, P. M. P. Piskunow, C. A. Balseiro, and G. Usaj, Phys. Rev. Lett. {\bf113}, 266801 (2014).

\bibitem{M.A.Sentef} M. A. Sentef, M. Claassen, A. F. Kemper, B. Moritz, T. Oka, J. K. Freericks, and T. P. Devereaux, Nat. Comm. {\bf6}, 7047 (2015).

\bibitem{A.Quelle} A. Quelle, M. Goerbig, C. M. Smith, arxiv: 1503.02635v2.

\bibitem{Perez2014} P. M. Perez-Piskunow, G. Usaj, C. A. Balseiro, and L. E. F. Foa Torres, Phys. Rev. B {\bf 89}, 121401(R) (2014).

\bibitem{Perez2015} P. M. Perez-Piskunow, L. E. F. Foa Torres, and G. Usaj, Phys. Rev. A {\bf 91}, 043625 (2015).


\bibitem{J.H.Shirley} J. H. Shirley, Phys. Rev. B {\bf138}, B979 (1965).

\bibitem{H.Sambe} H. Sambe, Phys. Rev. A {\bf 7}, 2203 (1973).

\bibitem{T.Oka} T. Oka and H. Aoki, Phys. Rev. B {\bf79}, 081406 (2009).

\bibitem{Wang2} Y. X. Wang, F. X. Li, and Y. M. Wu, EPL (Europhysics Letters) {\bf 105}, 17002 (2014).

\bibitem{M.Ezawa} M. Ezawa, Europhys. Lett. {\bf104}, 27006 (2013).

\bibitem{Y.X.Wang} Y. X. Wang, F. X. Li, and Y. M. Wu, EPL (Europhysics Letters) {\bf99}, 47007 (2012).

\bibitem{M.Lababidi} M. Lababidi, I. I. Satija, and E. H. Zhao, Phys. Rev. Lett. {\bf112}, 026805 (2014).

\bibitem{Z.Y.Zhou} Z. Y. Zhou, I. I. Satija, and E. H. Zhao, Phys. Rev. B {\bf90}, 205108 (2014).

\bibitem{A.Quelleb} A. Quelle and C. M. Smith, Phys. Rev. B {\bf90}, 195137 (2014). 

\bibitem{G.Jotzu} G. Jotzu, M. Messer, R. Desbuquois, M. Lebrat, T. Uehlinger, D. Greif, and T. Esslinger, Nature {\bf515}, 237 (2014).

\bibitem{P.Hauke} P. Hauke, O. Tieleman, A. Celi, C. Olschlager, J. Simonet, J. Struck, M. Weinberg, P. Windpassinger, K. Sengstock, M. Lewenstein, and A. Eckardt, Phys. Rev. Lett. {\bf109}, 145301 (2012).

\bibitem{M.Bellec} M. Bellec, U. Kuhl, G. Montambaus, and F. Mortessagne, Phys. Rev. B {\bf88}, 115437 (2013).

\bibitem{S.Gehler} S. Gehler, T. Tudorovskiy, C. Schindler, U. Kuhl and H. J. Stockmann, New J. Phys. {\bf15}, 083030 (2013).

\bibitem{F.Li} F. Li, L. Sheng,  and D. Y. Xing, EPL(Europhysics Letters) {\bf 86}, 60004(2009).

\bibitem{F.Li2} F. Li, L. B. Shao, L. Sheng, and D. Y. Xing, Phys. Rev. A {\bf 78}, 053617(2009).

\bibitem{J.T.Stewart} J. T. Stewart, J. P. Gaebler, and D. S. Jin, Nature (London) {\bf454}, 744 (2008).

\end{references}
\end{document}